\documentclass[11pt]{amsart}
\pdfoutput=1
\usepackage{geometry}                
\usepackage{float}
\usepackage{caption}
\usepackage{subcaption}
\usepackage[parfill]{parskip}    
\usepackage{graphicx}
\usepackage{amssymb}
\usepackage{epstopdf}
\usepackage[usenames,dvipsnames]{xcolor}
\usepackage{array}

\title{New Scaling Relation for Information Transfer in Biological Networks}

\author{Hyunju Kim$^1$} 
\author{Paul Davies$^1$}
\author{Sara Imari Walker$^{1,2}$}

\begin{document}
\maketitle
\vspace{-0.4in}
\begin{center}
{\tiny $^1$BEYOND: Center for Fundamental Concepts in Science, Arizona State University, Tempe, AZ}\\
{\tiny $^2$School of Earth and Space Exploration, Arizona State University, Tempe, AZ}
\end{center}
\begin{abstract}
Living systems are often described utilizing informational analogies. An important open question is whether information is merely a useful conceptual metaphor, or intrinsic to the operation of biological systems. To address this question, we provide a rigorous case study of the informational architecture of two representative biological networks: the Boolean network model for the cell-cycle regulatory network of the fission yeast {\it S. pombe} \cite{Davidich:2008cv} and that of the budding yeast {\it S. cerevisiae} \cite{Lietal:2004}. We compare our results for these biological networks to the same analysis performed on ensembles of two different types of random networks. We show that both biological networks share features in common that are not shared by either ensemble. In particular, the biological networks in our study, on average, process more information than the random networks. They also exhibit a scaling relation in information transferred between nodes that distinguishes them from either ensemble: even when compared to the ensemble of random networks that shares important topological properties, such as a scale-free structure. We show that the most biologically distinct regime of this scaling relation is associated with the dynamics and function of the biological networks. Information processing in biological networks is therefore interpreted as an {\it emergent property of topology} ({\it causal structure}) {\it and dynamics} ({\it function}). These results demonstrate quantitatively how the informational architecture of biologically evolved networks can distinguish them from other classes of network architecture that do not share the same informational properties.  
\end{abstract}

\section{Introduction}
Living systems are often described in terms of logic modules, information flows and computation \cite{Nurse2008Nature}. Such informational language is utilized in fields as diverse as evolutionary biology \cite{Goldenfeld2010}, neuroscience \cite{Lizier2011JComNeuro}, pattern formation \cite{Levin2014MolBioCell}, colonial decision making in eusocial insects \cite{Waters2012PlosOne}, and protein-protein interaction networks \cite{Bray1995Nature}, to name just a few. Although informational analogies are widely applied, an important open question is whether information is {\it intrinsic} to the operation of biological systems or merely a useful conceptual metaphor \cite{sep-information-biological, Griffiths2015PhilSci}. The debate over the ontological status of information in biology has far-reaching consequences, including implications for our understanding of whether biological organization is fully reducible to the known laws of physics and chemistry or new ``informational laws'' necessitating a foundational status for information in physical theory are necessary to account for living matter \cite{Marletto2015Interface, Deutsch2014RSA, Danchin2009FEMS, Auletta2008Interface}. The resolution to this debate should inform both our understanding of life's emergence, and what we should be looking for in the search for signatures of life beyond Earth. 

A weaker, and perhaps more widely accepted, perspective holds that while information is certainly useful in describing biological systems, ultimately all of biological complexity is, at least in principle, fully reducible to known physics. Under this view, although biological systems may appear complex, no new {\it physical principles} are needed to explain the phenomenon of life. By contrast, the stronger viewpoint takes the perspective that information is not merely descriptive, but instead is intrinsic to the operation of living systems \cite{Ellis2012Interface}. If the strong viewpoint holds, life would necessarily be classified as distinct from other kinds of physical systems, as we know of no other class of physical system where information is necessary to specify its state \cite{WD:2013} \footnote{With the exception of the trivial sense in which ``information'' is used to describe every physical system.}.  
A convincing case for the strong viewpoint must {\it quantitatively} satisfy two conditions: 

	(1) biological systems must be demonstrated to somehow be unique in their informational architecture, as compared to other classes of physical systems; and 
	(2) information must be shown necessary to the execution of biological function -- that is to say, information must be shown to matter to matter. 

The necessity of the first condition is perhaps obvious: if information is fundamental to biological organization, a strong signal should appear when contrasting information in living systems with the same concept of information as applied to non-living systems. The challenge is to define the appropriate concept of ``information''  in this context. The second criteria is less immediately obvious, but is clarified by a simple example. An informational pattern, such as the sequence of bases necessary to specify the GNRA tetraloop ``GAGA'', is readily copyable to other states of matter, {\it e.g.} such as this page of text. This kind of pattern cannot therefore be unique to life in the sense of the strong view, as other states of matter can share the same informational pattern. This may be one reason why attempts to quantify biological complexity in terms of Shannon Entropy have been relatively unsuccessful (for example, it is well known that genome size, which can be correlated with Shannon Information content, does not readily map to organismal complexity \cite{Eddy2012CurrBio}). What, if anything, characterizes life as unique state of matter quantified by ``information'' must therefore necessarily tie information to doing something, {\it e.g. to the causal structure underlying function} \cite{Griffiths2015PhilSci, Auletta2008Interface, Ellis2012Interface, davies2004arxiv}. It is for this reason that we utilize the terminology ``informational architecture'' rather than ``information'' or ``informational pattern''  herein, as architecture implies physical structure whereas mere patterns are not necessarily tied to their physical instantiation. Thus far, attempts to address the strong view have been primarily qualitative.

In what follows, we utilize simple, well-studied Boolean network models for real biological systems and demonstrate quantitatively that the biological networks studied have distinct informational architecture as compared to their random network counterparts. We do so by utilizing information theoretic analyses, detailing how information is processed in the execution of biological function. Our approach aims in part to address a research program laid out in a seminal paper by Nurse, wherein he calls for more focused understanding on {\it flows of information} within biological networks \cite{Nurse2008Nature}. We therefore focus on the concept of ``information flows'', interpreted as information processing and measured by Schreiber's transfer entropy \cite{Schreiber2000PRL} as a concept of information that allows us to address conditions 1 and 2 above. {\it We conjecture that if the strong viewpoint holds, it is how living systems implement informational correlations (process information) through space and time in the execution of function that sets them apart from other classes of physical systems. }We note that while there has been a great deal of emphasis placed on understanding the ``logic of life'', quantitative results have thus far been primarily topological {\it or} dynamical. Examples include analysis of the topology of network motifs -- or logic circuits -- necessary to biological function \cite{Milo2002Science, Alon2006}, or dynamical network features such as the robustness of the global attractor landscape \cite{Huang2005PRL}. The approach we present herein is distinct from these previous efforts in that we explicitly address information processed, which, as we will show, should be viewed as an emergent property of networks that arises from the integration of topology {\it and} dynamics.

Herein, we focus on two model systems developed previously: the Boolean network model for the cell-cycle regulatory network of the fission yeast 
({\it S. pombe}) \cite{Davidich:2008cv} and that of the budding yeast 
({\it S. cerevisiae}) \cite{Lietal:2004}. We calculate the information transferred between pairs of nodes within each network in the execution of function and contrast the results with the same analysis performed on random networks of two different classes: Erd\"{o}s-R\'{e}nyi (ER) networks and Scale-Free (SF) networks. The latter share a scale-free structure with the cell-cycle networks, which is often cited as a distinguishing feature of many real-world networks including biological networks \cite{Albert2002RevMoPhy, Caldarelli2007OUP, Eguiluz2005PRL, Albert2005JCellSci}. We show that both cell-cycle networks share commonalities in their informational architecture that set them far apart from their random network counterparts. One of the most striking features uncovered is a scaling relation for the distribution of information transfer between nodes, which for the cell-cycle networks is statistically different from that observed for either ensemble of random networks, despite the topological resemblance between the cell-cycle networks and SF networks in our study. We identified the regime where the cell-cycle networks differ most significantly from the random networks in their scaling relation, and characterized the patterns in information transfer relative to the function (dynamics) of each network. The results show that control kernel nodes, which have previously been connected with regulation of function, play an important role in information transfer within the cell-cycle networks. We also investigated how the causal structure (topology) of the cell-cycle networks affects information transfer as compared to the random networks and found that the scale-free structure, shared by the cell-cycle and SF networks, utilizes long-range correlations more than direct causal interactions between nodes for information processing, unlike what is observed for ER networks. These results are suggestive of previously unidentified information-based organizational principles that go beyond topological considerations, which may be critical to biological function. Finally, we found that both cell-cycle networks process more information than the majority of random networks in either ensemble, suggestive of evolutionary optimization for information processing. 

We note that our analysis, being based on standard information measures, is not level specific, and thus the approach is expected to generalize to most, if not all, biological networks. Therefore, we potentially offer an operational means to quantify life or to detect generic signatures of life, in terms of its informational properties. The results presented in the paper thus open a new framework for addressing the debate over the status of information in biology, by demonstrating {\it quantitatively} how the informational architecture of biologically evolved networks distinguishes them from other classes of network architecture that do not share the same informational properties. 

\section{Model and Methods} 
The study presented herein is a first attempt to identify patterns in information processing that might be distinctive of biological organization, as compared to other classes of physical systems, and in turn to connect these patterns to causal structure (topology) and function (dynamics). Our analysis requires an integrative synthesis of a number of distinct areas, including: Boolean network models of biological function, information theoretic analysis for distributed computation, sampling networks with topological constraints, and control of cellular behavior. We briefly describe each herein. 

\subsection{Boolean Network Models for the Cell-Cycle Regulatory Process} \label{subset: cell-cycle net}
\begin{figure}[H]
\begin{center}
\includegraphics[width=6.5cm]{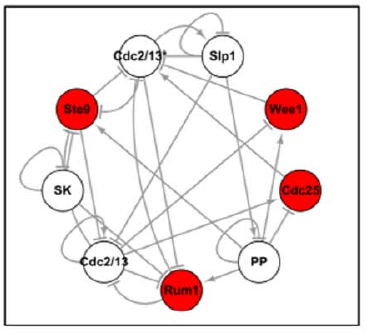}\hspace{1.0cm}
\includegraphics[width=6.5cm]{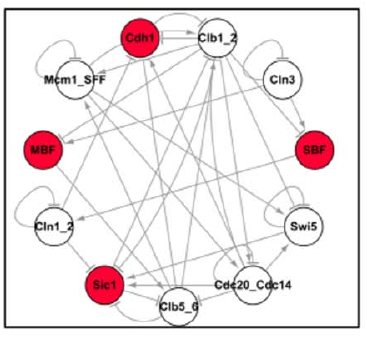}
\caption{Boolean network models of fission (left) and budding (right) yeast cell-cycle regulation. Nodes represent the regulatory proteins and edges denote two types of biochemical interactions between nodes: activation (ended with an arrow) and inhibition (ended with a bar). The nodes colored red are the {\it control kernel}, which regulates the global behavior of each network when pinned to specific values. Figure adopted from \cite{Kim2013SciRep}}
\label{fig:net-ck}
\end{center}
\end{figure}

Boolean network models have proven in many cases to provide accurate models for biological function \cite{Albert2003JTheoBio, Espinosa2004Plant, Mendoza1999Bioinfo}.
They are also the most readily tractable network models for information-theoretic analysis, since each node may take on only one of two discrete states \cite{Lizieretal2008, Lizier2011ALife}. They are thus ideal for our case study. In this study we focus on the cell-cycle regulatory networks of the fission yeast {\it S. pombe} \cite{Davidich:2008cv} and of the budding yeast {\it S. cerevisiae} \cite{Lietal:2004}. The Boolean network models for both systems are shown in Fig. \ref{fig:net-ck}. Each node corresponds to one protein among the small subset of key regulatory proteins involved in each respective cell-cycle. The state of node $i$, $S_i (t) \in \{0,1\}$, indicates whether the given protein is present (1) or not (0). Biochemical causal interactions between proteins are denoted by edges in the network. Time advances through the cycle and the states of nodes are updated (in parallel) in discrete time steps according to the following rule:

\begin{eqnarray} \label{eq:updating}
    S_i (t +1)=\left\{
                \begin{array}{ll}
                  1, \qquad \quad \sum_{j} a_{ij}S_j (t) > \theta_i .\\
                  0, \qquad \quad \sum_{j} a_{ij}S_j (t) < \theta_i .\\
                  S_i (t), \quad \;\, \sum_{j} a_{ij}S_j (t) = \theta_i .
                \end{array}
              \right.
 \end{eqnarray}
 
\noindent where $a_{ij}$ is the edge weight between node $i$ and $j$ ($a_{ij} = -1$ for inhibition links and  $a_{ij} = 1$ for activation links), and $\theta_i$ is the threshold for node $i$. The threshold for all nodes in both networks is $0$, with the exception of Cdc2/13 and Cdc2/13*, which have thresholds of $-0.5$ and $0.5$, respectively. 

Although numerous Boolean network models for biological systems exist, we specifically focus on these two networks because they are small and accurately model biological function. The small network size, with $\sim 10$ nodes each, permits statistically reliable comparison of results for the biological networks to the average properties of randomized networks of the same size (number of nodes and edges), due to the relatively small ensemble size of comparable random networks. Thus, for these two networks, we can readily address condition (1) as posed above, by making a meaningful comparison between each biological network and an ensemble of random networks with similar size and topological features. Connecting information processing and function, as laid out by criterion (2), is also tenable for both models. For both cell-cycle networks, there is a direct connection between the dynamics on these networks and the corresponding biological function: both networks correctly reproduce the sequences of protein states corresponding to the phases of the respective cell-cycle (see \cite{Davidich:2008cv, Lietal:2004} for details). Therefore, any distinctive patterns uncovered in our analysis of informational architecture can be related to  dynamics, and consequently, the biological function of each network. 

We note in particular that the task of addressing condition (2) is in general not a trivial one. Our approach is to connect information processing to topology {\it and} dynamics, that is to the causal mechanisms of each biological network that define its function \cite{Balduzzi2008PlosComBio} (as we will show in addressing condition (1), topology alone is not sufficient to quantify information processing in the biological networks). Stated differently, condition (2) requires identifying how information transfer might depend on causal structure in non-trivial ways such that {\it information processing is intrinsic to function} (and therefore matters to matter). We note that the edges connecting nodes for both cell-cycle networks are modeled based on experimental data detecting a direct causal interaction between the proteins they represent \cite{Davidich:2008cv, Lietal:2004}. Comparing information transfer between nodes connected by an edge to that between nodes that are not directly interacting therefore provides insights into how correlations are distributed amongst nodes in the execution of function. 

Our analysis focuses on a few key nodes of each cell cycle network, called the {\it control kernel}, that regulate the dynamics of the entire network. These are highlighted in red in Fig. \ref{fig:net-ck}. Control kernels were recently discovered by Kim {\it et al} in a number of biological networks, and seem to be a generic feature of Boolean models for regulatory networks \cite{Kim2013SciRep}. A control kernel is defined as the minimal set of nodes such that pinning their value to that of the primary attractor associated with biological function guarantees the convergence of the network state to that attractor. This property led Kim {\it et al} to conclude that the control kernel acts as a local mechanism for regulating the global behavior the network. To address condition (2) we study how information transfer is related to the causal mechanisms of both cell cycles by determining the distribution of information transfer among pairs of nodes with a causal connection and without, and more specifically we analyze information transfer through control kernel nodes that regulate function.

We note that despite the fact that budding and fission yeast are closely related genetically \cite{Tyson2001Nature}, closely related genes between the two organisms can play vastly different functional roles \cite{Forsburg1999TrenGen}. Also, while the two networks share similar overall dynamics in terms of the dominating largest attractor \cite{Davidich:2008cv, Lietal:2004}, they show significant differences in their underlying biochemical machinery \cite{Simanis2003JCellSci}. Thus, for the purposes of our study, we view them as two independent examples of biological networks with related function. Studying both networks in parallel provides a comparative analysis to look for features common to biological networks that are not shared by their random counterparts.

\subsection{Construction of Random Boolean Networks with Biological Constraints} \label{subsec: random} 
To identify any features distinctive to biologically evolved networks that might satisfy criteria (1) or (2), we compare the results of our information--theoretic analysis on both cell-cycle networks to the same analysis performed on ensembles of random networks. Scale-free structure is ubiquitous in real-world networks from the Internet to social and biochemical networks. Scale free structure occurs in cases where there exist hub nodes with significantly more connections than others, and where the distribution of the number of edges for each node follows a power-law function over the whole network. Many features of biological networks (e.g. robustness to random failure) \cite{Albert2002RevMoPhy}, as well as others classified as scale-free networks, have been explained as arising from the scale-free structure. Here, we are in particular interested in going beyond a scale-free structure and addressing whether information processing (which we attribute to topology and dynamics) is a more distinctive feature of biological networks than global topology (scale-free structure) alone. We therefore compare our results for the biologically functional cell-cycle networks to two different classes of random networks that act as controls for analyzing properties distinctive of biological networks: Erd\"{o}s-R\'{e}nyi (ER) networks and Scale-Free (SF) networks. 

For meaningful comparison, both classes of random networks were constructed under constraints with reference to each cell-cycle network (see Table \ref{tab:random}). ER network here indicates networks sampled networks using an Erd\"{o}s-R\'{e}nyi random graph model, where every pair of nodes is connected to an edge with a fixed probability \cite{Erdos1959PublMath, Erdos1960PubMathHun}. Therefore, we use the term ``ER networks" to distinguish them from another class of randomly sampled network (SF) utilized in this study. We note our ER networks are equivalent to the class of networks commonly referred to as ``random networks" in other literature. In our study, the term Scale-Free network, unlike its common definition, does not mean that sample networks in the ensemble exhibit power-law degree distributions. Due to their small size, even the degree distributions of the biological cell-cycle networks do not follow a power-law. Instead, the term, ``scale-free" as used in this paper, emphasizes that the sample networks have the {\it same exact degree sequence}, defined as the number of edges for each node over the whole network as each cell-cycle network (see Table \ref{tab:random}), and hence the networks share the same bias in terms of global structure as the cell-cycle networks. In this paper, the SF networks are generated by edge-swapping from the reference cell-cycle network \cite{Maslov2002Science} (for a more general method, see \cite{HJK2009, HJK2010, HJK2012}). We note that since having the same degree sequence is a sufficient condition for sharing a degree distribution, the analogous randomly generated networks with reference to larger biological networks that are truly scale-free would also be scale-free. Therefore, the comparison of the cell-cycle network to both class of random networks allows us to isolate the contribution attributable to global topological features such as degree distribution, which the biological networks share with the SF networks but not the ER networks in our sample, from any informational structure that may arise solely due to network architecture which is peculiar to biological function.

\subsection{Quantifying Information Processing}\label{subset:info-te} 
Our information-theoretic analysis focuses on quantifying ``information processing'' in the biological and random networks in our study.  Our motivation is to capture Nurse's notion of ``information flows''\footnote{Herein we use the terms ``flow'', ``transfer'' and ``processing'' interchangeably in reference to information, and quantify this concept of information using Schreiber's transfer entropy \cite{Schreiber2000PRL}. This is more informal language than the technical meaning of ``information flow'' as formulated by Ay and Polani, which is a measure of causal flows \cite{Ay2008}. Since we do not implement the Ay and Polani measure herein we use ``flow'' to directly connect our quantitative results with the notion of ``information flow'' as introduced by Nurse (which does not necessarily imply direct causal interaction).} as a concept of information relevant to biological function \cite{Nurse2008Nature}. We adopt the notion of information processing as implemented in information dynamics, a formalism for quantifying the component operations of computation in dynamical systems, utilizing the tools of information theory. In information dynamics, information processing is quantified using Schreiber's transfer entropy \cite{Schreiber2000PRL, Lizieretal2008, Lizier2010EurPhys, Wangetal2011}, a directional measure of information transfer. Transfer entropy (TE) from a source node $Y$ to a target node $X$ is defined as the reduction in uncertainty due to knowledge of state of $Y$ about the future state of $X$, above the reduction in uncertainty provided by knowledge about the past states of $X$. TE from $Y$ to $X$ can be written as:

\begin{eqnarray} \label{eq:TE}
T_{Y \rightarrow X} (k) = \sum \limits_{(x_{n}^{(k)}, x_{n+1}, y_n) \in \mathcal{A}_0}  p(x_{n}^{(k)}, x_{n+1}, y_n) \log_2 \frac{p(x_{n+1} | x_{n}^{(k)}, y_n)}{p(x_{n+1} | x_n^{(k)})} ~.
\end{eqnarray}
where $ \mathcal{A}_0$ indicates the set of all possible patterns of sets of states $(x_{n}^{(k)}, x_{n+1}, y_n)$ and $x_n^{(k)}$ denotes $(x_n, . . . , x_{n-k+1})$, the vector of $k$ previous states of destination $X$ at time-step $n+1$. Also, $y_n$ and $x_{n+1}$ represent the state of $Y$ at the current time step and the  state of $X$ at the next time step. The probability distributions in Eq. \ref{eq:TE} are defined as the relative frequency of each pattern of states over the times series of dynamical states of the network. In our study, to obtain the probability distributions for each cell cycle network and the random network ensembles, we generated every possible trajectory for each network by applying the updating rule in Eq. \ref{eq:updating} up through 20 time steps  for all the $2^n$ possible initial network states (i.e. all possible combinations of binary states for nodes in the network), where $n$ is the total number of nodes in the particular network under study. We chose 20 time steps as the length of the trajectory generated from each initial state, since it is sufficiently long to capture transient dynamics for networks before converging on a fixed point (an attractor) for the cell-cycle and for the vast majority of random networks. 

From Eq. \ref{eq:TE}, one can see that TE is the mutual information between $y_n$ and $x_{n+1}$ conditioned on $k$ previous states of $X$. Directionality arises due to the asymmetry between time steps for the state of the source and the destination in Eq \ref{eq:TE}. Due to this asymmetry, TE can be utilized to measure ``information flows''. It is therefore more appropriate for our purposes of quantifying distinct features of information processing in biological networks than related measures such as mutual information, which measure correlations only, without reference to directionality. We note that TE is correlational and does not necessarily reflect direct causal interactions between pairs of nodes, but may take on non-zero values even for pairs of nodes that do not directly interact \cite{Lizier2010EurPhys}. The causal structure for the networks studied herein is fully described by their edges as noted in Section \ref{subset: cell-cycle net}.

\section{Results}
\subsection{Scaling Relation for Information Transfer in Yeast Cell-Cycle Networks}  \label{subsec:scale}
\begin{figure}[H]
\begin{center}
\begin{subfigure}[b]{0.79\textwidth}
\includegraphics[width=\textwidth]{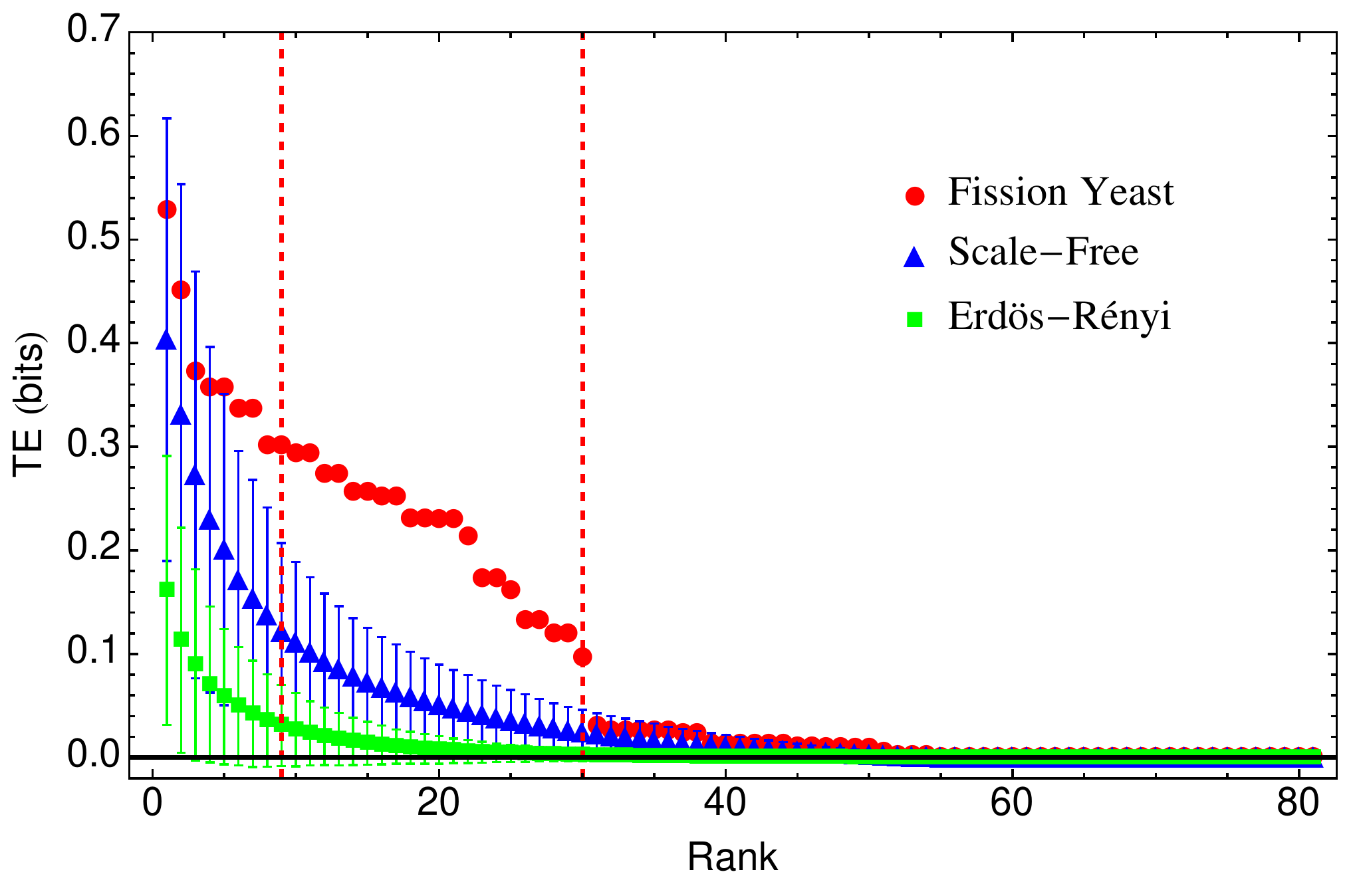}
\caption{Fission Yeast}
\end{subfigure}
\begin{subfigure}[b]{0.8\textwidth}
\includegraphics[width=\textwidth]{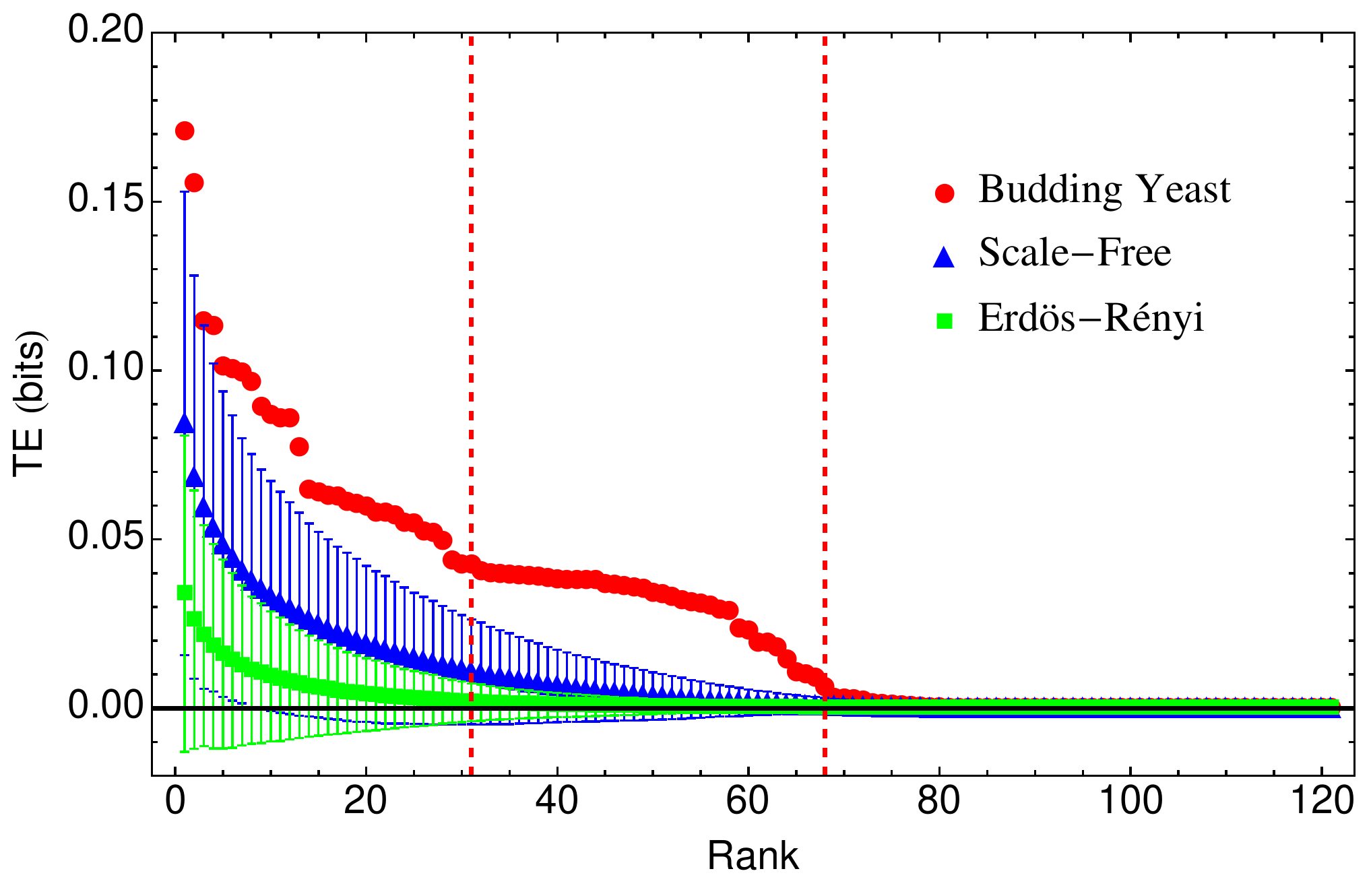}
\caption{Budding Yeast}
\end{subfigure}
\caption{Scaling of information processing (TE) among pairs of nodes for cell-cycle (red), ER (green) and SF (blue) networks. History lengths for computing TE were $k = 2$ and $k=5$ for the fission and the budding yeast, respectively. The averages and the standard deviation of for each of the random networks ensembles are computed for a sample of $1000$ networks. Regions between dashed lines denote $\chi$ for each cell-cycle network. }
\label{fig:te-scale}
\end{center}
\end{figure}

To reiterate, we use transfer entropy (TE) as a candidate measure for identifying features potentially distinctive to biological networks, focusing our analysis on the concept of ``information flows'' that may be particular to biological function as suggested by Nurse \cite{Nurse2008Nature}. For every ordered pair of nodes $(i, j)$ in both cell cycle networks, we calculated TE from node $i$ to node $j$, $T_{i \rightarrow j}$ (as described in Section \ref{subset:info-te}) and ranked the pair $(i,j)$ according to its measured value of TE. The same analysis was performed on each network in the ensembles of ER and SF networks as a point for comparison to identify any features particular to the biological networks. The resulting scaling relations are shown in Fig. \ref{fig:te-scale}, where biological networks are highlighted in red for fission yeast (top) and budding yeast (bottom). The ensemble averages for the scaling relations of ER and SF random networks generated with reference to each cell cycle are shown in green and blue, respectively, in each respective panel. Error bars represent the standard deviation over the ensembles of random networks (averages are over $1,000$ random networks). For the results shown in Fig. \ref{fig:te-scale} the history lengths $k=2$ and $k=5$ were selected since these history lengths show the most distinctive scaling distribution as compared to the two random network ensembles for the fission and budding yeast networks, respectively (see Supplement for scaling relations over history lengths $k = 1, \dots, 8$). The scaling distributions reveal a non-linear relationship between the information transferred between pairs of nodes (y-axis) and their relative rank (x-axis), for each of the network classes studied -- biological, SF and ER. The scaling relation is most striking for the biological networks (red), which are significant outliers with respect to either of the ER or SF ensembles. 

Fig. \ref{fig:te-scale} shows that ER networks have much less information transfer than SF or biological networks. This result suggests that scale-free structure -- as is the case for the SF and biological networks, but not the ER networks -- plays as significant role in increasing information transfer within a network. The deviation between ER networks and SF networks or the cell-cycle networks may be expected due to the differences in topological features, which the transfer entropy is partially based on. However, surprisingly, scale-free structure alone does not account for the high level of information transfer between nodes in the biological networks. The biological networks differ from the SF networks despite their common topological features. With the exception of the few highest ranked node pairs, the biological networks exhibit information transfer that is several standard deviations larger than the corresponding rank in the SF ensemble for the majority of ranks with $TE \neq 0$. The excess TE observed in the biological networks in Fig. \ref{fig:te-scale} deviates between $1\sigma$ to $5\sigma$ from that of the SF networks, with a trend of increasing divergence from the SF ensemble for lower ranked node pairs that still exhibit correlations ({\it e.g.}, where $TE > 0$). We define $\chi$ as the set of node pairs whose TE deviates $> 2\sigma$ from the SF networks, indexed by rank. The ranks that deviate more than $2\sigma$ for the fission yeast are $\chi_f = \{9,10 \ldots 30\}$ and for budding yeast $\chi_b = \{31, 32 \ldots 68\}$ (highlighted between the dashed lines in each panel in Fig \ref{fig:te-scale}). Patterns in the biological distribution also exhibit plateaus, suggestive that there exist subgroups wherein informational flow is evenly distributed, as opposed to a few dominating informational connections. We analyze this substructure in Sections \ref{subsec:ck-scale} in terms of the dynamics (function) of the cell cycle networks.

\subsection{Total Information Processed}
\begin{figure}[H]
\begin{center}
\begin{subfigure}[b]{0.79\textwidth}
\includegraphics[width=\textwidth]{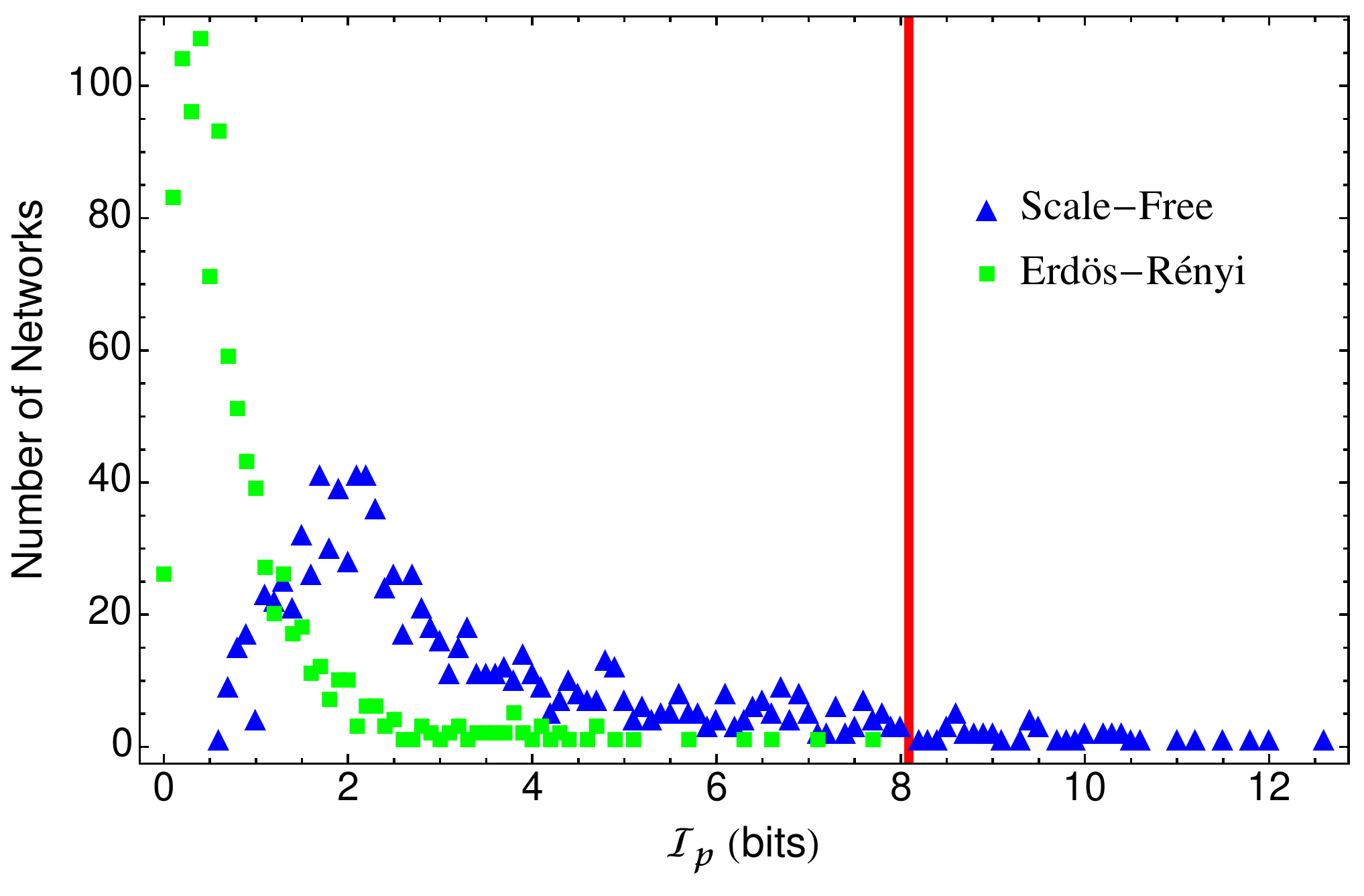} 
\caption{Fission Yeast}
\end{subfigure}
\begin{subfigure}[b]{0.8\textwidth}
\includegraphics[width=\textwidth]{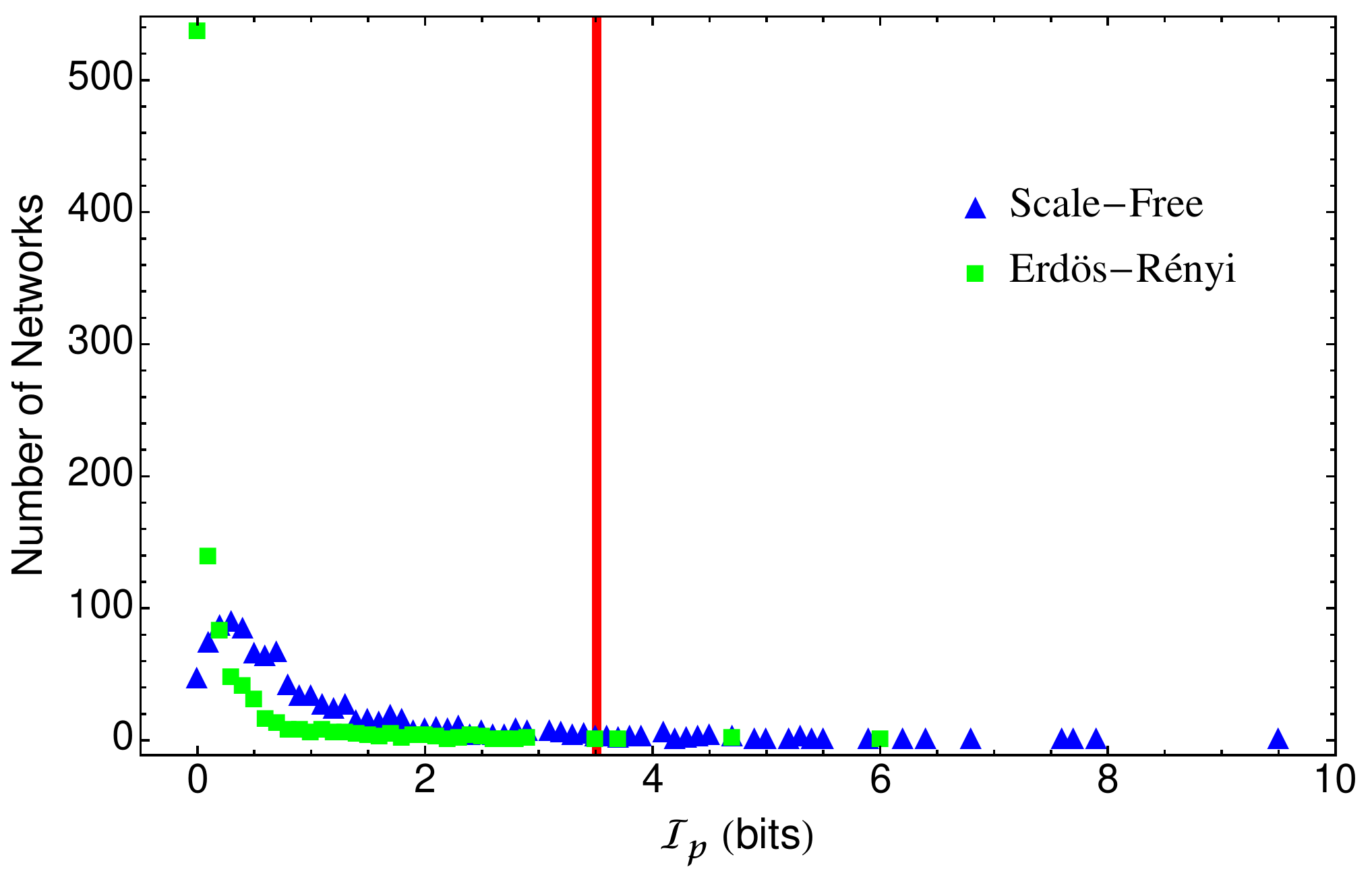}
\caption{Budding Yeast}
\end{subfigure}
\caption{Distributions for total information processed (sum of TE for all pairs of nodes) for the ensembles of ER (green) and SF (blue) networks.  Each data point represents the number (y-axis) of individual networks within the respective ensemble with a given amount of total information transferred (x-axis). Also red line indicates the total information processed for the fission (Upper) and budding (Lower) yeast cell-cycle network. }
\label{fig:Ip_freq}
\end{center}
\end{figure}

We also calculated the total information processed by the cell-cycle networks and compared to that of individual instances of SF and ER networks within the random ensembles. We define a quantity called {\it total information processed}, denoted by $I_p$, which is the sum of the transfer entropy between all pairs of nodes $(i,j)$ for an individual network, $I_p = \sum_{(i,j)} T_{i \rightarrow j} $, for a given history length $k$ ( $k=2$ and $k=5$ for the fission and the budding yeast, respectively). Accordingly, the total information processed by the fission and budding yeast cell-cycle networks is $I_p = 8.09$ and $I_p = 3.51$, respectively -- shown as red lines in the Fig. \ref{fig:Ip_freq}. The frequency distributions of networks associated with $I_p$ for the two sets of SF and ER networks are shown in Fig. \ref{fig:Ip_freq}. Comparing the two distributions, it is evident that, on average, SF networks process more information than ER networks, with higher frequencies for networks with larger values of $I_p$. Fig. \ref{fig:Ip_freq} also shows that the biological networks process more information than most random networks in the ensembles. More specifically, the fission yeast cell-cycle network lies outside of $95\%$ of the SF networks and $100\%$ of the ER networks. Similarly, the budding yeast cell-cycle network is in the $95.1\%$ percentile of SF networks and the $99.5\%$ percentile of ER networks. This result indicates that the cell-cycle networks are highly optimized, with only $1/2000$ ER networks and $1/200$ SF networks displaying comparable total information processed in a random draw. 

\subsection{Distribution of Information Processing over Causal Structure} \label{subsec:causal}
The correlations measured with TE can arise due to direct causal effect ({\it e.g.} via an edge) or statistical correlations between two nodes that are not directly connected via a causal interaction ({\it e.g.} via long range correlations). We note that since we are using transfer entropy in our analysis, the measured correlations are across {\it space} (between nodes) and {\it time} (between discrete time steps)-- they therefore differ from correlations typically associated with critical phenomena as applied to networks, which are usually strictly spatial ({\it e.g.} such as mutual information). 

To identify whether the distinctive features of the cell-cycle networks shown in Fig. \ref{fig:te-scale} and Fig. \ref{fig:Ip_freq} arise due to information transfer along edges, or longer-range correlations, we classified each pair of nodes as either having a direct causal interaction (connected by an edge) or not (no edge). We also classified them as being correlated ($TE > 0$) or not ($TE =0$). The results of this classification scheme are shown in Table \ref{tab:edge-te}, and are very similar for both cell cycle networks. In both cases, the majority ($> 40\%$) of node pairs are correlated via information transfer, even though they are not causally connected by an edge. Roughly $25\%$ of node pairs have a causal interaction with information transfer, or no interaction and no information transfer. The remaining minority of nodes exhibit a causal interaction with no corresponding transfer of information (causation without correlation). The same analysis was applied to SF and ER networks corresponding to each cell-cycle, shown in Fig. \ref{fig:edge-te-ran} (see Supplement for detailed data). Fig. \ref{fig:edge-te-ran} shows a distinctive transition in the distribution of correlations moving from the ER to SF networks ensembles. In the ER networks, the majority of node pairs that are not connected by an edge are also not correlated. By contrast, the majority of node pairs in SF networks are correlated even though they are not directly connected, and this pattern is even more prominent for the biological cell-cycle networks. Biological networks therefore appear to be highly optimized for correlations among nodes, even in cases where there is no direct causal interaction. 

\subsection{Information Transfer and Regulation of Biological Function} \label{subsec:ck-scale}
We analyzed the scaling relations for the cell-cycle networks in Fig. \ref{fig:te-scale}, with particular focus on the biologically distinct regime $\chi$, in terms of the global causal structure of each biological network. While the local causal structure of these networks are fully articulated by the edges shown in Fig. \ref{fig:net-ck}, we view the global causal structure as embedded in the relationship between control kernel nodes and the flow of dynamics to the associated attractors in network state space, as discussed in Section \ref{subset: cell-cycle net} (see also discussion in \cite{Walker2015PhilA}). 


Here we are interested in how the control kernel nodes -- as drivers of global dynamics and functional regulators -- play a role in information transfer. To do so, we divided all nodes in each cell-cycle network into two groups: CK and NCK, where CK denotes the set of control kernel nodes identified in \cite{Kim2013SciRep} and NCK is the complement of CK, i.e. non-control kernel nodes. Therefore, an ordered pair of nodes $(i, j)$ in the networks falls in one of four groups shown in Table \ref{inter-te} depending on whether  $i$ or $j$ belongs to CK or NCK. We specified the groups for each node pair in the scaling patterns for information transfer for both biological networks (red in Fig. \ref{fig:te-scale}), as shown in Fig. \ref{fig:te-causal-de}. 

The highest ranked regimes of the cell cycle scaling relations shown in Fig. \ref{fig:te-scale}, where the biological networks differ least from the scale-free networks, are dominated by information transfer between NCK nodes (NCK $\rightarrow$ NCK, shown in aqua), see  Fig. \ref{fig:te-causal-de}. The most biologically distinctive regime ($\chi_f$ and $\chi_b$) is, by contrast, dominated by information transfer between CK nodes and NCK nodes, i.e. CK $\rightarrow$ NCK (purple) and NCK $\rightarrow$ CK (orange) (see Supplement Table 1). For the biological networks the scaling regime that deviates most from the SF networks is dominated by information transfer through the control kernel nodes. This suggests that for both cell cycle networks, the patterns in information processing observed to be most distinct to the biological networks are strongly affected by the presence of the control kernel, and hence must be associated with the regulation of function.

\section{Conclusions}
It is an open question whether the concept of information as applied to biological systems is merely a useful conceptual metaphor or hints at deeper physical principles underlying biological organization. Support for the viewpoint that information is not merely descriptive, but instead integral to the operation of living systems (strong view) requires that at least two conditions are satisfied: (1) that biological systems must be demonstrated to somehow be unique in their informational architecture, as compared to other classes of physical systems; and, (2) that information must be shown to be intrinsic to operation of biological systems, e.g. to the execution of biological function. Our analyses presented here provide first quantitative data addressing these conditions, with results lending support to the view that biological systems are distinctive in their informational architecture.

Our results indicate that scale-free structure -- characterizing the biological and SF networks in our study, but not the ER networks -- plays as significant role in increasing information transfer within a network. This result is particularly interesting when considered within the context of the wide-spread observations of scale-free structure in various biological networks \cite{Albert2002RevMoPhy, Caldarelli2007OUP, Eguiluz2005PRL, Albert2005JCellSci}. The ubiquity of scale-free networks is prima facie explained as a result of their robustness properties. However, the high levels of information transfer in SF networks as compared to ER networks uncovered in this study indicates an alternative explanatory hypothesis may also hold true -- that is, that SF networks arise as a result of their amplified information processing. 

Significantly, the enhancement of SF networks over ER networks is not sufficient to explain the information transfer observed for the biological networks. Condition (1) is therefore  satisfied by the statistically significant difference between ``information flows" of biological networks and that of ensembles of ER and SF networks: the scale-free structure taken alone is not sufficient to explain the distribution of information processing for either biological network. This result has important implications for modeling of biological function, which has thus far primarily focused on topological {\it or} dynamic properties. In particular, topological features of scale-free networks -- such as power-law degree distribution -- are often viewed as sufficient to capture essential features of biological organization due precisely to the fact that they are observed in a number of disparate biological phenomena  \cite{Albert2002RevMoPhy, Albert2005JCellSci}. Our results indicate that an additional criterion for accurately modeling biological function, is that network models include the dynamics of information transfer in their construction, and in particular that models should optimize information transfer between nodes, since this feature distinguishes the biological networks in our study from generic SF networks. {\it The newly discovered informational scaling relation is an emergent property of networks that arises from the integration of topology and dynamics, which cannot be accounted for solely by one of these two features.}

Additionally, the biological networks in our study are outliers in terms of the total information processed, as quantified by $I_p$, in each network in our study. This concept of information (information processing) satisfies the condition of being not readily copyable to other states of matter, since it is the physical instantiation (causal mechanisms) that give rise to the observed patterns in information flows, and thus is a candidate for satisfying Condition (2).  Our analysis also in part directly addresses condition (2), by connecting the observed distinctive features of the informational architecture of biological networks in Fig. 2 to their causal structure and their function. The enhancement of information transfer between nodes for biological networks, as compared to SF or ER, is primarily due to correlations between nodes which are not directly causally connected (non-local) -- a frequent signature of collective, or critical states of organization. Typically, criticality is described in terms of long-range correlations in space. Herein, we have shown that for the cell cycle networks, correlations are in {\it space and time}, i.e., are associated with information processing. Interestingly, both networks studied have very similar patterns for the distribution of TE among causally connected node pairs (Table 2). It is an open question whether this pattern is indicative of cell-cycle function, or a more general pattern of biological organization that might be characteristic of networks with other functions. In particular, since the cell cycle is optimized for {\it processing information through time} as a mechanism for keeping track of phases during cellular division, a question of interest is whether other networks, optimized for {\it processing spatial information} (for example, the genetic regulation of embryo development \cite{Davidson1998Sci}), exhibit the same or different informational patterns in their distribution of correlations among causal edges.

An important feature of both networks in this study, which is shared with other regulatory networks, is the control kernel, defined as a subset of nodes that regulate the attractor dynamics associated with biological function. Our analysis has revealed that most of the biologically distinct regimes of the information transfer scaling relations, $\chi_f$ and $\chi_b$ for the fission and budding yeast respectively, is attributable to the presence of the control kernel. Information transfer in or out of the control kernel dominates the biologically distinct regime for both networks. Furthermore, for the budding yeast network, the ranks deviating most from random are primarily those corresponding to information transfer between control kernel nodes. Although not conclusive that information is intrinsic to function (e.g. that condition (2) holds), our results clearly indicate that the patterns in information processing unique to the biological networks in our study are attributable to the regulation of function by a few key nodes. Interestingly, these nodes have other properties consistent with their information theoretic interpretation: in particular, the control kernel provides a mechanism for {\it distinguishability} among attractor states for the biological networks. As noted by Kim et al., the set of control kernel nodes takes on a unique and distinct state in every attractor state in the networks studied \cite{Kim2013SciRep}. They thus provide a means for coarse-graining the state space in a functionally relevant manner (such that the primary attractor associated with function is distinguishable from other attractor states). That the network organizes information flows through these nodes as it executes dynamics of the cell cycle is a highly non-trivial feature of both networks' informational structure. We also note that the control kernel nodes were effectively discovered by a causal intervention in the manner of Pearl \cite{Pearl2000}, i.e. by fixing the subset of nodes to their value in the biologically relevant attractor. Combined with our results, this suggests interesting connections between information transfer and top-down causal regulation \cite{Walker2014} of biological function, discussed in more depth in \cite{Walker2015PhilA}.

We hypothesize that the features reported herein may be common to biological networks of different function, and in particular, that scaling relations in information transfer may be a hallmark of biological organization. Our results are suggestive of previously unidentified information-based organizational principles that go beyond topological considerations such as scale-free structure, and may be critical to biological function. They thus open a new framework for addressing the debate over the status of information in biology, by demonstrating quantitatively how the informational architecture of biologically evolved networks can distinguish them from other classes of network architecture that do not exhibit the same informational properties as reported here.
 
\section{Acknowledgements}
This project/publication was made possible through support of a grant from Templeton World Charity Foundation. The opinions expressed in this publication are those of the author(s) and do not necessarily reflect the views of Templeton World Charity Foundation. The authors wish to thank Larissa Albantakis for thoughtful suggestions on constructing random graphs and Joseph Lizier for constructive conversations on generating the probability distributions necessary for the information-theoretic analysis. 

\bibliographystyle{unsrt}
\bibliography{scalingTE-arXiv}

\begin{thebibliography}{10}

\bibitem{Davidich:2008cv}
Maria~I Davidich and Stefan Bornholdt.
\newblock {Boolean Network Model Predicts Cell Cycle Sequence of Fission
  Yeast}.
\newblock {\em PloS ONE}, 3(2):e1672, February 2008.

\bibitem{Lietal:2004}
Fangting Li, Tao Long, Ying Lu, Qi~Ouyang, and Chao Tang.
\newblock {The yeast cell-cycle network is robustly designed}.
\newblock {\em PNAS}, 2004.

\bibitem{Nurse2008Nature}
Paul Nurse.
\newblock Life, logic and information.
\newblock {\em Nature}, 454(7203):424--426, 2008.
\newblock 10.1038/454424a.

\bibitem{Goldenfeld2010}
Nigel Goldenfeld and Carl Woese.
\newblock Life is physics: evolution as a collective phenomenon far from
  equilibrium.
\newblock {\em arXiv:1011.4125}, 2010.

\bibitem{Lizier2011JComNeuro}
Joseph~T Lizier, Jakob Heinzle, Annette Horstmann, John-Dylan Haynes, and
  Mikhail Prokopenko.
\newblock Multivariate information-theoretic measures reveal directed
  information structure and task relevant changes in fmri connectivity.
\newblock {\em Journal of Computational Neuroscience}, 30(1):85--107, 2011.

\bibitem{Levin2014MolBioCell}
Michael Levin.
\newblock Molecular bioelectricity: how endogenous voltage potentials control
  cell behavior and instruct pattern regulation in vivo.
\newblock {\em Molecular biology of the cell}, 25(24):3835--3850, 2014.

\bibitem{Waters2012PlosOne}
James~S. Waters and Jennifer~H. Fewell.
\newblock Information processing in social insect networks.
\newblock {\em PLoS ONE}, 7(7):e40337, 2012.

\bibitem{Bray1995Nature}
Dennis Bray.
\newblock Protein molecules as computational elements in living cells.
\newblock {\em Nature}, 376(6538):307--312, 1995.

\bibitem{sep-information-biological}
Peter Godfrey-Smith and Kim Sterelny.
\newblock Biological information.
\newblock In Edward~N. Zalta, editor, {\em The Stanford Encyclopedia of
  Philosophy}. Fall 2008 edition, 2008.

\bibitem{Griffiths2015PhilSci}
Paul~E. Griffiths, Arnaud Pocheville, Brett Calcott, Karola Stotz, Hyunju Kim,
  and Rob Knight.
\newblock Measuring causal specificity.
\newblock {\em Philosophy of Science}, To be appeared, 2015.

\bibitem{Marletto2015Interface}
Chiara Marletto.
\newblock Constructor theory of life.
\newblock {\em Journal of The Royal Society Interface}, 12(104):20141226, 2015.

\bibitem{Deutsch2014RSA}
David Deutsch and Chiara Marletto.
\newblock Constructor theory of information.
\newblock {\em Proceedings of the Royal Society of London A: Mathematical,
  Physical and Engineering Sciences}, 471(2174), 2014.

\bibitem{Danchin2009FEMS}
Antoine Danchin.
\newblock Bacteria as computers making computers.
\newblock {\em FEMS microbiology reviews}, 33(1):3--26, 2009.

\bibitem{Auletta2008Interface}
Gennaro Auletta, G.~F.~R. Ellis, and L.~Jaeger.
\newblock Top-down causation by information control: from a philosophical
  problem to a scientific research programme.
\newblock {\em Journal of The Royal Society Interface}, 5(27):1159--1172, 2008.

\bibitem{Ellis2012Interface}
George F.~R. Ellis, Denis Noble, and Timothy O'Connor.
\newblock {\em Top-down causation: an integrating theme within and across the
  sciences?}, volume~2.
\newblock 2012.
\newblock Journal Article 2012-02-06 00:00:00 Interface Focus 1.

\bibitem{WD:2013}
Sara~I. Walker and Paul Davies.
\newblock {The algorithmic origins of life}.
\newblock {\em J. Roy. Soc. Interface}, 6:20120869, 2013.

\bibitem{Eddy2012CurrBio}
Sean~R. Eddy.
\newblock The c-value paradox, junk dna and encode.
\newblock {\em Current biology}, 22(21):R898--R899, 2012.

\bibitem{davies2004arxiv}
Paul C.~W. Davies.
\newblock Emergent biological principles and the computational properties of
  the universe.
\newblock {\em arXiv preprint astro-ph/0408014}, 2004.

\bibitem{Schreiber2000PRL}
Thomas Schreiber.
\newblock Measuring information transfer.
\newblock {\em Physical Review Letters}, 85(2):461--464, 2000.
\newblock PRL.

\bibitem{Milo2002Science}
Ron Milo, Shai Shen-Orr, Shalev Itzkovitz, Nadav Kashtan, Dmitri Chklovskii,
  and Uri Alon.
\newblock Network motifs: simple building blocks of complex networks.
\newblock {\em Science}, 298(5594):824--827, 2002.

\bibitem{Alon2006}
Uri Alon.
\newblock {\em An introduction to systems biology: design principles of
  biological circuits}.
\newblock CRC press, 2006.

\bibitem{Huang2005PRL}
Sui Huang, Gabriel Eichler, Yaneer Bar-Yam, and Donald~E. Ingber.
\newblock Cell fates as high-dimensional attractor states of a complex gene
  regulatory network.
\newblock {\em Physical review letters}, 94(12):128701, 2005.

\bibitem{Albert2002RevMoPhy}
Albert-L{\'a}szl{\'o} Barab{\'a}si and R{\'e}ka Albert.
\newblock Statistical mechanics of complex networks.
\newblock {\em Reviews of modern physics}, 74(1):47, 2002.

\bibitem{Caldarelli2007OUP}
Guido Caldarelli.
\newblock Scale-free networks: complex webs in nature and technology.
\newblock {\em OUP Catalogue}, 2007.

\bibitem{Eguiluz2005PRL}
Victor~M. Eguíluz, Dante~R. Chialvo, Guillermo~A. Cecchi, Marwan Baliki, and
  A.~Vania Apkarian.
\newblock Scale-free brain functional networks.
\newblock {\em Physical Review Letters}, 94(1):018102, 2005.
\newblock PRL.

\bibitem{Albert2005JCellSci}
R{\'e}ka Albert.
\newblock Scale-free networks in cell biology.
\newblock {\em Journal of cell science}, 118(21):4947--4957, 2005.

\bibitem{Albert2003JTheoBio}
R{\'e}ka Albert and Hans~G Othmer.
\newblock The topology of the regulatory interactions predicts the expression
  pattern of the segment polarity genes in drosophila melanogaster.
\newblock {\em Journal of Theoretical Biology}, 223(1):1--18, 2003.

\bibitem{Espinosa2004Plant}
Carlos Espinosa-Soto, Pablo Padilla-Longoria, and Elena~R. Alvarez-Buylla.
\newblock A gene regulatory network model for cell-fate determination during
  arabidopsis thaliana flower development that is robust and recovers
  experimental gene expression profiles.
\newblock {\em The Plant Cell}, 16(11):2923--2939, 2004.

\bibitem{Mendoza1999Bioinfo}
Luis Mendoza, Denis Thieffry, and Elena~R Alvarez-Buylla.
\newblock Genetic control of flower morphogenesis in arabidopsis thaliana: a
  logical analysis.
\newblock {\em Bioinformatics}, 15:593--606, 1999.

\bibitem{Lizieretal2008}
Joseph~T Lizier, Mikhail Prokopenko, and Albert~Y Zomaya.
\newblock The information dynamics of phase transitions in random boolean
  networks.
\newblock In {\em ALIFE}, pages 374--381, 2008.

\bibitem{Lizier2011ALife}
Joseph~T Lizier, Siddharth Pritam, and Mikhail Prokopenko.
\newblock Information dynamics in small-world boolean networks.
\newblock {\em Artificial Life}, 17(4):293--314, 2011.

\bibitem{Balduzzi2008PlosComBio}
David Balduzzi and Giulio Tononi.
\newblock Integrated information in discrete dynamical systems: motivation and
  theoretical framework.
\newblock {\em PLoS Comput Biol}, 4(6):e1000091, 2008.

\bibitem{Kim2013SciRep}
Junil Kim, Sang-Min Park, and Kwang-Hyun Cho.
\newblock Discovery of a kernel for controlling biomolecular regulatory
  networks.
\newblock {\em Scientific reports}, 3, 2013.

\bibitem{Tyson2001Nature}
John~J. Tyson, Kathy Chen, and Bela Novak.
\newblock Network dynamics and cell physiology.
\newblock {\em Nat Rev Mol Cell Biol}, 2(12):908--916, 2001.
\newblock 10.1038/35103078.

\bibitem{Forsburg1999TrenGen}
Susan~L. Forsburg.
\newblock The best yeast?
\newblock {\em Trends in Genetics}, 15(9):340--344, 1999.

\bibitem{Simanis2003JCellSci}
Viesturs Simanis.
\newblock Events at the end of mitosis in the budding and fission yeasts.
\newblock {\em Journal of cell science}, 116(21):4263--4275, 2003.

\bibitem{Erdos1959PublMath}
Paul Erd{\"o}s and Alfr{\'e}d R{\'e}nyi.
\newblock On random graphs. i.
\newblock {\em Publ. Math. Debrecen}, 6:290--297, 1959.

\bibitem{Erdos1960PubMathHun}
Paul Erd{\"o}s and Alfr{\'e}d R{\'e}nyi.
\newblock On the evolution of random graphs.
\newblock {\em Publ. Math. Inst. Hung. Acad. Sci}, 5:17--61, 1960.

\bibitem{Maslov2002Science}
Sergei Maslov and Kim Sneppen.
\newblock Specificity and stability in topology of protein networks.
\newblock {\em Science}, 296(5569):910--913, 2002.

\bibitem{HJK2009}
Hyunju Kim, Zolt{\'a}n Toroczkai, P{\'e}ter~L Erd{\H{o}}s, Istv{\'a}n
  Mikl{\'o}s, and L{\'a}szl{\'o}~A Sz{\'e}kely.
\newblock Degree-based graph construction.
\newblock {\em Journal of Physics A: Mathematical and Theoretical},
  42(39):392001, 2009.

\bibitem{HJK2010}
Charo~I Del~Genio, Hyunju Kim, Zolt{\'a}n Toroczkai, and Kevin~E Bassler.
\newblock Efficient and exact sampling of simple graphs with given arbitrary
  degree sequence.
\newblock {\em PloS one}, 5(4):e10012, 2010.

\bibitem{HJK2012}
Hyunju Kim, Charo~I Del~Genio, Kevin~E Bassler, and Zolt{\'a}n Toroczkai.
\newblock Constructing and sampling directed graphs with given degree
  sequences.
\newblock {\em New Journal of Physics}, 14(2):023012, 2012.

\bibitem{Ay2008}
Nihat Ay and Daniel Polani.
\newblock Information flows in causal networks.
\newblock {\em Advances in Complex Systems}, 11(1):17--41, 2008.

\bibitem{Lizier2010EurPhys}
Joseph~T Lizier and Mikhail Prokopenko.
\newblock Differentiating information transfer and causal effect.
\newblock {\em The European Physical Journal B}, 73(4):605--615, 2010.

\bibitem{Wangetal2011}
X~Rosalind Wang, Jennifer~M Miller, Joseph~T Lizier, Mikhail Prokopenko, and
  Louis~F Rossi.
\newblock Measuring information storage and transfer in swarms.
\newblock In {\em Proceedings of the Eleventh European Conference on the
  Synthesis and Simulation of Living Systems}, pages 838--845, 2011.

\bibitem{Walker2015PhilA}
Sara~I. Walker, Hyunju Kim, and Paul Davies.
\newblock The informational architecture of the cell.
\newblock {\em arXiv:1507.03877}, 2015.

\bibitem{Davidson1998Sci}
Chiou-Hwa Yuh, Hamid Bolouri, and Eric~H. Davidson.
\newblock Genomic cis-regulatory logic: Experimental and computational analysis
  of a sea urchin gene.
\newblock {\em Science}, 279(5358):1896--1902, 1998.

\bibitem{Pearl2000}
Judea Pearl.
\newblock {\em Causality: models, reasoning and inference}, volume~29.
\newblock MIT press Cambridge, 2000.

\bibitem{Walker2014}
Sara~I. Walker.
\newblock Top down causation and the rise of information in the emergence of
  life.
\newblock {\em Information}, 5:424--439, 2014.

\end{thebibliography}

\newpage
\begin{table} 
\begin{tabular}{| >{\centering\arraybackslash}m{1in} || >{\centering\arraybackslash}m{2.1in} | >{\centering\arraybackslash}m{2.1in} |}
	\hline
	 & {Erd\"{o}s-R\'{e}nyi (ER) networks}&{Scale-Free (SF) networks}\\
	\hline\hline
		{Size of network (Total number of nodes, inhibition and activation links)} & Same as the cell-cycle network & Same as the cell-cycle network  \\
		\hline
		{Nodes with a self-loop}&Same as the cell-cycle network &Same as the cell-cycle network \\
		\hline
		{Threshold for each node}&Same as the cell-cycle network &Same as the cell-cycle network \\
		\hline
		{The number of activation and inhibition links for each node}& {\bf NOT} the same as the cell-cycle network ($\rightarrow$ no structural bias)  & Same as the cell-cycle network ( $\rightarrow$ scale-free structure)  \\
		  \hline
        \end{tabular} 
        \bigskip
        \caption{Constraints for constructing two different classes of random networks that retain features of the causal structure of a reference cell-cycle network.}
 \label{tab:random}        
\end{table}

\begin{table} 
    \begin{subtable}{.45\linewidth}
        \centering
   		\begin{tabular}{  |   l   |   l   |   l   |   }
		\hline
			& {Edge } &{No Edge} \\
		  \hline\hline
		TE $>$ 0 &  23.46\% &  43.21\%  \\
		\hline
      		TE = 0 & 7.41\% & 25.93\% \\
		\hline
		\end{tabular} 
		 \caption{Fission Yeast}	
    	\end{subtable}
\begin{subtable}{.45\linewidth}
       		 \centering
        		\begin{tabular}{ |   l   |   l   |   l   |    }
		\hline
			& {Edge} &{No Edge} \\
		  \hline\hline
		TE $>$ 0 &  22.31\% & 48.76\% \\
      		\hline
		TE = 0 &  5.79\% &  23.14\% \\
		\hline
        		\end{tabular}  
        		\caption{Budding Yeast}
    \end{subtable}
     \caption{The distribution of TE within the cell-cycle networks and corresponding SF and ER networks, classified by pairs of nodes that are correlated ($TE > 0$) or not ($TE=0$) and causally interacting (edge) or not (no edge). The values indicate the ratio of the number of node pairs in each category to the total number of node pairs for each cell-cycle network.} \label{tab:edge-te} 
\end{table}

\begin{table} 
\begin{tabular}{  |   l   ||   l   |    l   | }
	\hline
	 & \hspace{2cm}{CK}&\hspace{2cm}{NCK}\\
	\hline\hline
	\vspace{-0.1in}
		&&\\
		{CK} & \multicolumn{1}{m{5cm}|}{\includegraphics[scale=0.4]{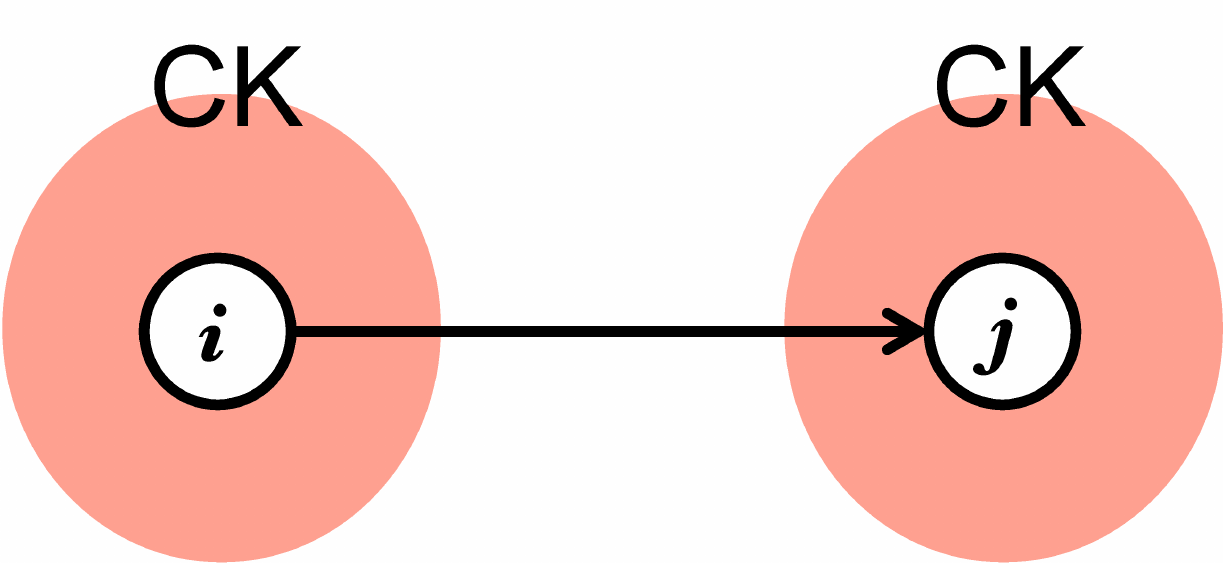}} &\multicolumn{1}{m{5cm}|}{\includegraphics[scale=0.4]{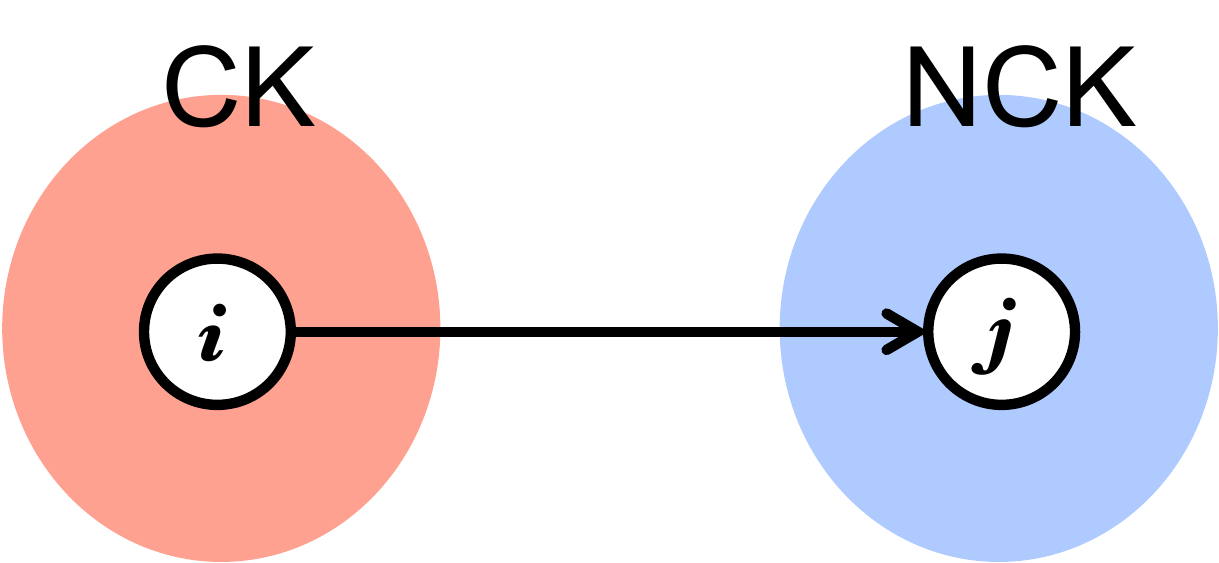}} \\
		\hline
		\vspace{-0.1in}
		&&\\
		{NCK}&\multicolumn{1}{m{5cm}|}{\includegraphics[scale=0.4]{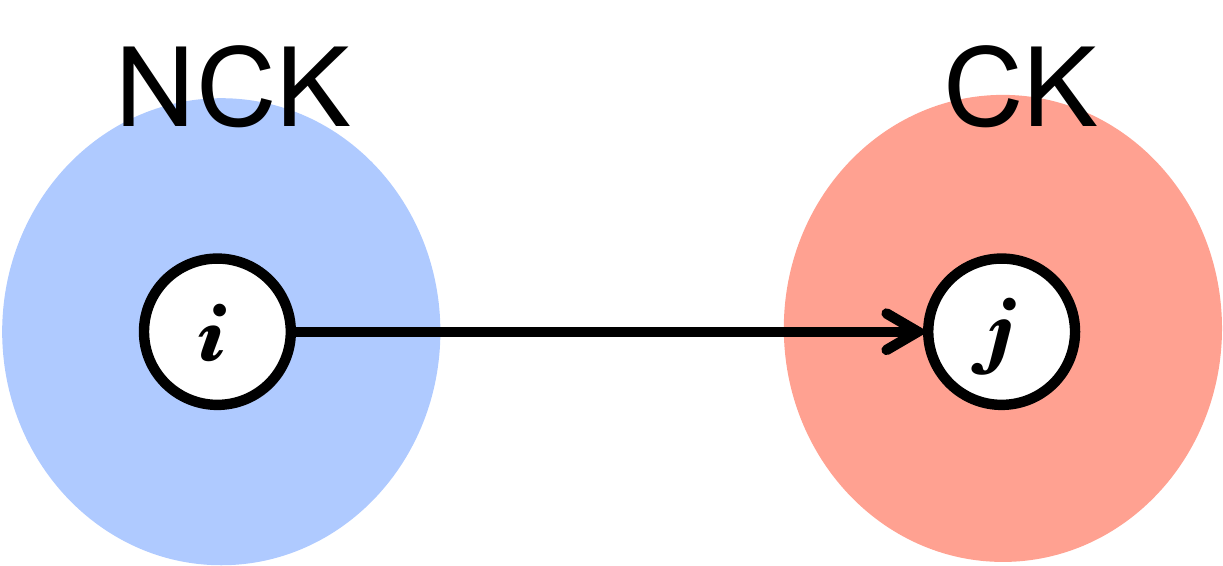}} & \multicolumn{1}{m{5cm}|}{\includegraphics[scale=0.4]{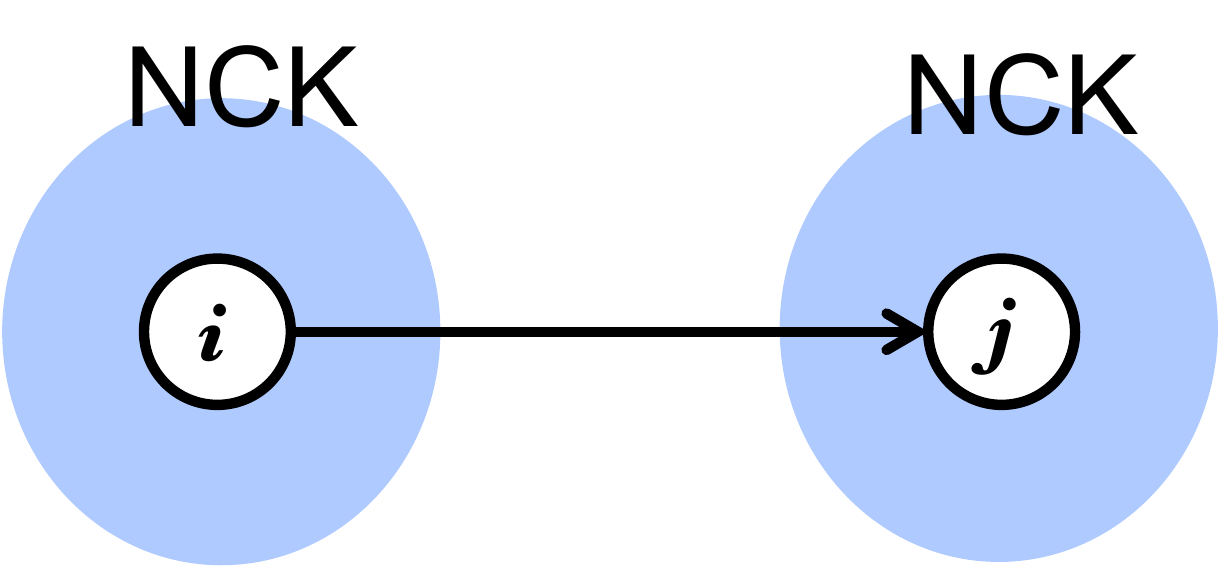}}  \\
		  \hline
        \end{tabular} 
        \bigskip
        \caption{Possible information flows between sets of nodes classified as control kernel (CK) (highlighted in red in Fig. \ref{fig:net-ck}) or non-control kernel (NCK). }
 \label{inter-te}        
\end{table}

\clearpage
\newpage

\begin{figure}[H]
\begin{center}
\begin{subfigure}[b]{0.8\textwidth}
\includegraphics[width=\textwidth]{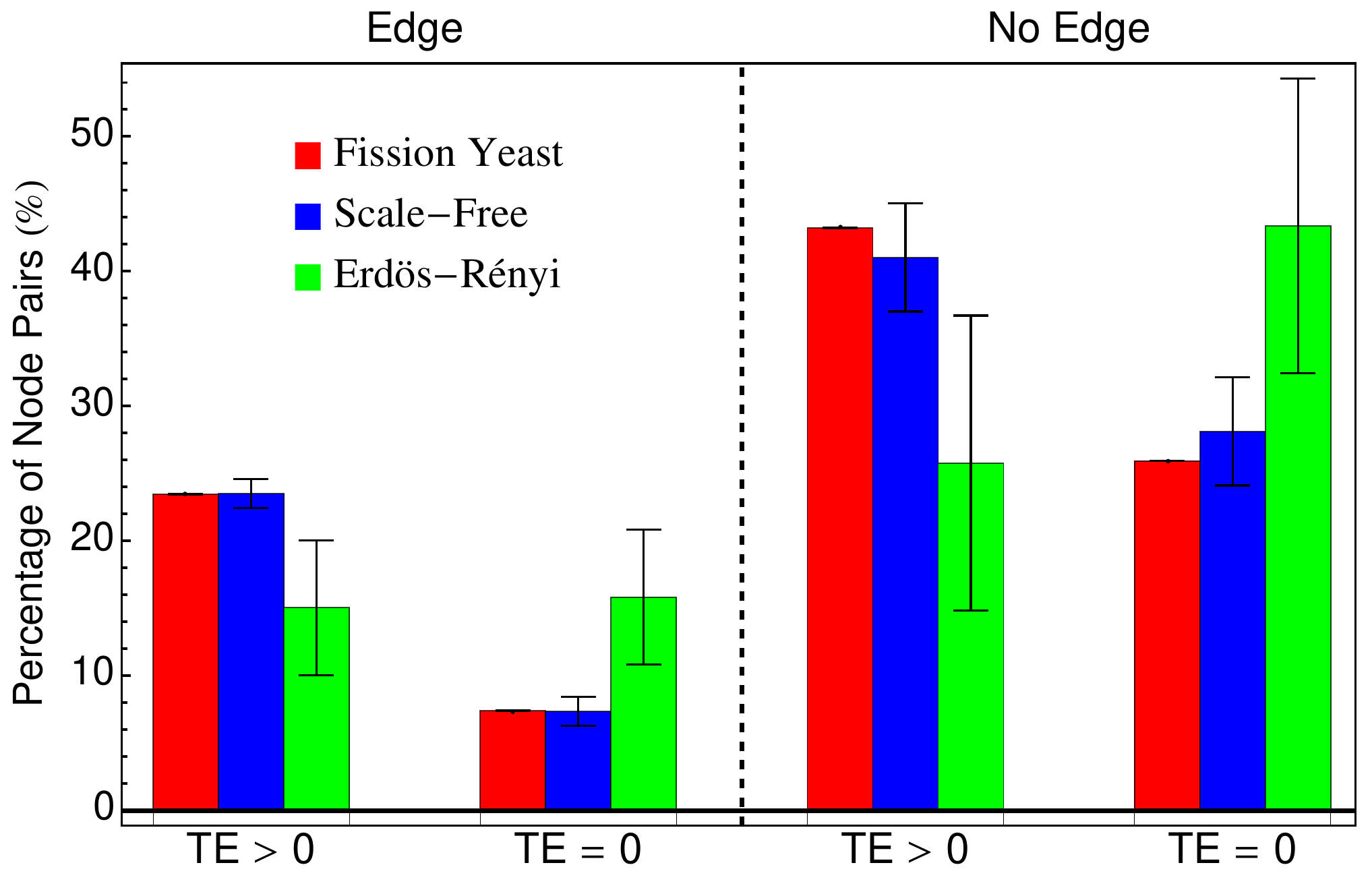}
\caption{Fission Yeast}
\end{subfigure}
\begin{subfigure}[b]{0.8\textwidth}
\includegraphics[width=\textwidth]{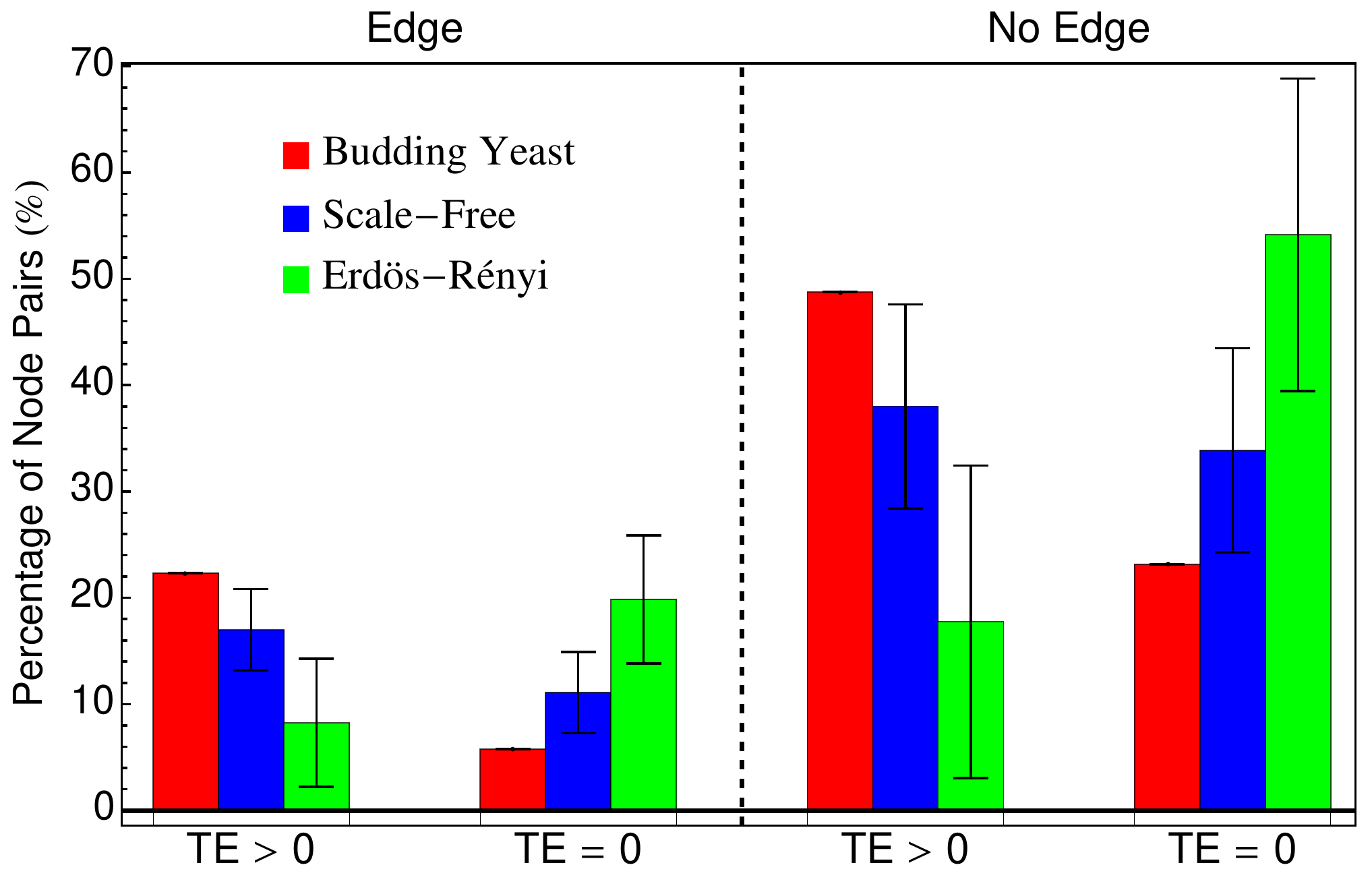}
\caption{Budding Yeast}
\end{subfigure}
\caption{Classification of all pairs of nodes within the cell-cycle (red), SF (blue) and ER (green) networks by correlation ($TE > 0$ or  $TE=0$) and causal interaction (edge or no edge). Each data bar indicates the percentage of node pairs (on Y-axis) within each category.}
\label{fig:edge-te-ran}
\end{center}
\end{figure}

\begin{figure}[H]
\begin{center}
\begin{subfigure}[b]{0.79\textwidth}
	\includegraphics[width=\textwidth]{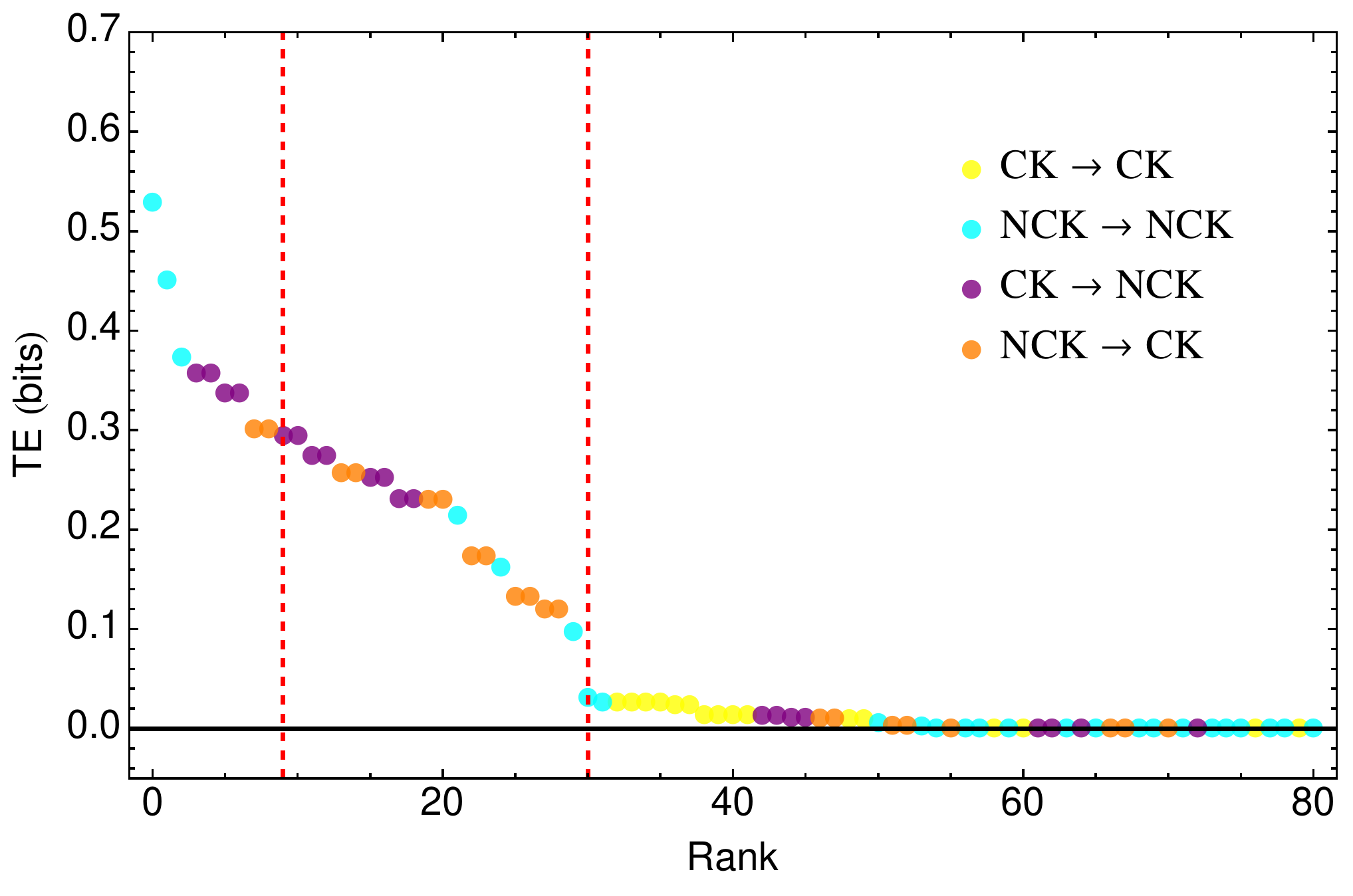}
	\caption{Fission Yeast}
\end{subfigure}
\begin{subfigure}[b]{0.8\textwidth}
	\includegraphics[width=\textwidth]{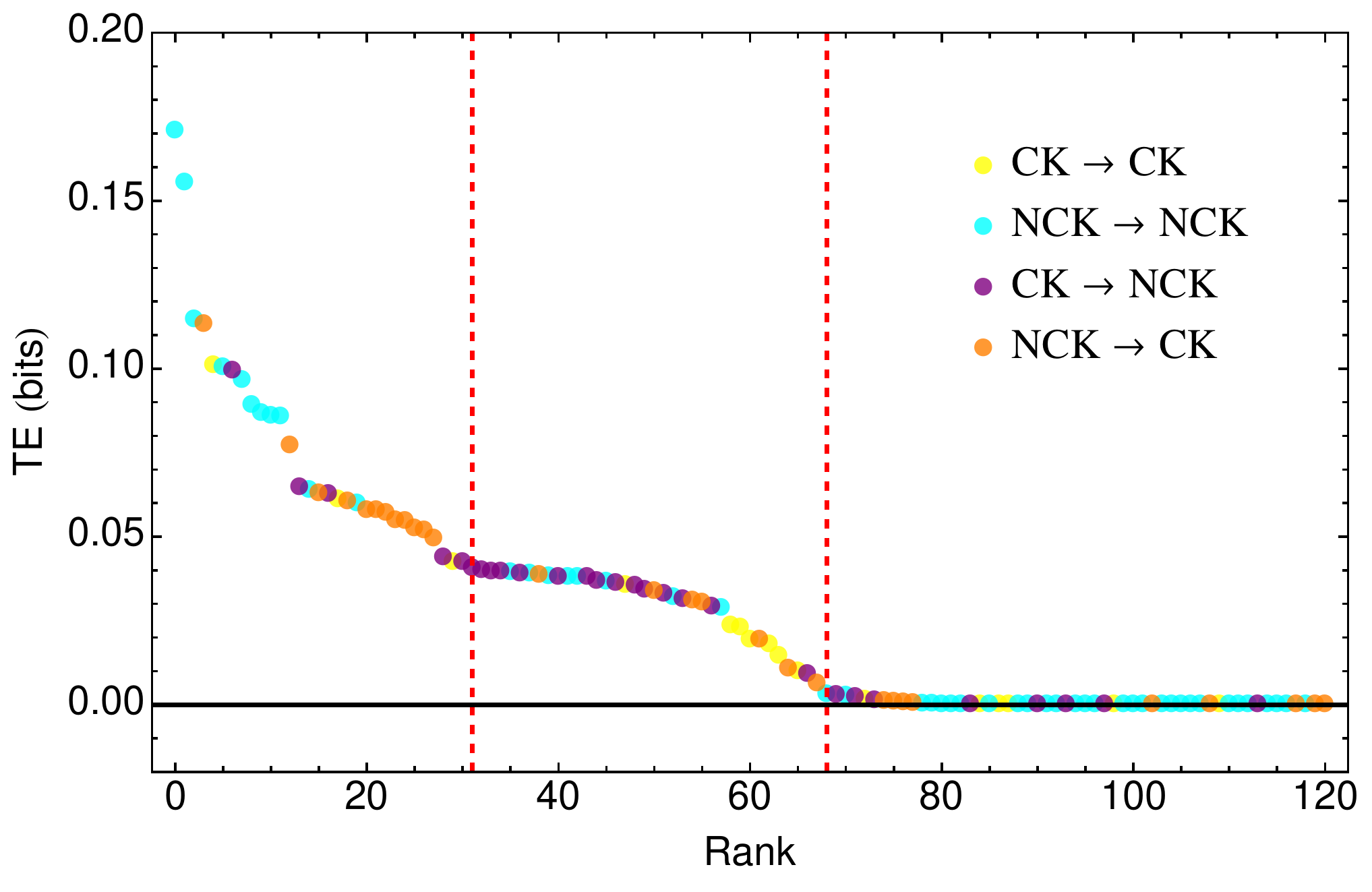}
	\caption{Budding Yeast}
\end{subfigure}
\caption{Scaling relations for information transfer for the fission yeast (top) and budding yeast (bottom) cell-cycle networks. Data shown is the same as in Fig. \ref{fig:te-scale} with data pointes divided into four classes of information transfer: $CK \rightarrow CK$ (yellow), $CK \rightarrow NCK$ (purple), $NCK \rightarrow CK$ (orange), and $NCK \rightarrow NCK$ (aqua).}
\label{fig:te-causal-de}
\end{center}
\end{figure}

\end{document}